\documentclass{article}
\usepackage{fullpage}
\usepackage{enumitem}
\usepackage{amsmath}

\usepackage{amsthm}
\usepackage{mathtools}
\usepackage[english]{babel}
\usepackage{amsxtra}
\usepackage{lipsum}
\usepackage{amsfonts}
\usepackage{graphicx}
\usepackage{subfig}
\usepackage{url}
\usepackage{epstopdf}
\usepackage{algorithmic}
\usepackage{caption}
\usepackage{float}
\newtheorem{theorem}{Theorem}

\DeclareMathOperator{\sign}{sign}
\DeclareMathOperator{\Diag}{Diag}

\newtheorem{myassump}{Assumption}
\def\Tr{\mathop{\rm \boldmath Tr}}

\newcommand\blfootnote[1]{%
	\begingroup
	\renewcommand\thefootnote{}\footnote{#1}%
	\addtocounter{footnote}{-1}%
	\endgroup
}

\begin{document}
%		\begin{center}
%		\footnotesize
%		\hrulefill
%		
%		\vspace{0.3em}
%		
%		© 2024 IEEE. Personal use of this material is permitted. 
%		The final version of this paper appeared in \textit{2024 32nd Telecommunications Forum (TELFOR)}.  
%		DOI:10.1109/TELFOR63250.2024.10819058
%		
%		\vspace{0.3em}
%		\hrulefill
%	\end{center}
	
	\begin{center}
		{\bf {\LARGE{Tackling heavy-tailed noise in distributed estimation: Asymptotic performance and tradeoffs }}}

		\vspace{1em}
		
				\large{
					\begin{tabular}{cc}
							Dragana Bajovic$^\dagger$,& Dusan Jakovetic$^\ast$,  
					\end{tabular}
					\begin{tabular}{ccc}
					 & Soummya Kar$^\rhd$, & Manojlo Vukovic$^\circ$ 
					\end{tabular}
				}
			
		\vspace*{.2in}
			
		\begin{tabular}{c}
			$^\dagger$ University of Novi Sad, Faculty of Technical Sciences, Department of \\Power, Electronic and
			Communication Engineering(dbajovic@uns.ac.rs) \\ 
		$^\ast$ University of Novi Sad, Faculty of Sciences, Department of\\ Mathematics and Informatics (dusan.jakovetic@dmi.uns.ac.rs)
		\\ 
		$^\rhd$ Department of Electrical and Computer Engineering, \\Carnegie Mellon University (soummyak@andrew.cmu.edu)
		\\
		$^\circ$University of Novi Sad, Faculty of Technical Sciences, Department of\\ Fundamental Sciences (manojlo.vukovic@uns.ac.rs)\\
		\end{tabular}
				
				\vspace*{.2in}
				
			%	\today
				
				\vspace*{.2in}
		
	\end{center}
	
	\begin{abstract}
	We present an algorithm for distributed estimation of an unknown vector parameter $\boldsymbol{\theta}^\ast \in {\mathbb R}^M$ in the presence of heavy-tailed observation and communication noises. Heavy-tailed noises frequently appear, e.g.,  
	in densely deployed Internet of Things (IoT) or wireless sensor network systems. The presented algorithm falls within the class of \emph{consensus+innovation} estimators and combats the effect of the heavy-tailed noises by adding general nonlinearities in the consensus and innovations update parts. We present results on almost sure convergence and asymptotic normality of the estimator. In addition, we provide novel analytical studies  that reveal interesting tradeoffs between the system noises and the underlying network topology.
	\blfootnote{
		\noindent \textbf{Acknowledgement. }The work of D. Bajovic and M. Vukovic is supported by the Ministry of Science, Technological Development and Innovation (Contract No. 451-03-65/2024-03/200156) and the Faculty of Technical Sciences, University of Novi Sad through project “Scientific and Artistic Research Work of Researchers in Teaching and Associate Positions at the Faculty of Technical Sciences, University of Novi Sad” (No. 01-3394/1). The work of D. Jakovetic is supported by the Ministry of Education, Science and Technological Development, Republic of Serbia. The work of D. Bajovic and M. Vukovic is also supported by the Serbian Ministry of Science, Technological development and Innovation within the bilateral project Serbia-Slovakia No. 337-00-3/2024-05/16. The work of M. Vukovic is also supported by the Science Fund of the Republic of Serbia, GRANT No 7359, Project title- LASCADO. The work of D. Jakovetic is also supported by Provincial Secretariat for Higher Education and Scientific Research, grant no. 142-451-2593/2021-01/2. The work of D. Jakovetic and D. Bajovic is also supported by the European Union’s Horizon Europe program under grant agreement number 101093006.
	}
	\blfootnote{© 2024 IEEE. Personal use of this material is permitted. 
		The final version of this paper appeared in \textit{2024 32nd Telecommunications Forum (TELFOR)}.
		DOI:10.1109/TELFOR63250.2024.10819058}
	\end{abstract}
	
\section{Introduction}
\label{section-intro}
This paper is concerned with \emph{consensus+innovations} distributed estimation in networked systems, e.g.,~\cite{KMR,SoummyaAdaptive}, when the inter-neighbor communication and network nodes' observations are contaminated with  noise that is heavy-tailed and can have an infinite variance. Heavy-tailed  noise appears, e.g., due to interference arising from neighboring devices in dense IoT or wireless sensor network deployments~\cite{Haenggi,IoTHeavyTail}.

In more detail, we consider zero-mean communication and sensing noises that may have infinite moments of order $\alpha,$ for any $\alpha>1.$ 
%To the best of our knowledge, prior works on \emph{consensus+innovations} distributed estimation and related distributed estimation schemes, e.g.,~\cite{KMR,SoummyaAdaptive}, always assume a finite-variance noise. Actually, 
It can been shown that, in the presence of heavy-tailed noise, standard \emph{linear}  \emph{consensus+innovations} estimators can fail to converge (see~\cite{Ourwork,Ourwork1}). In this paper, we present a \emph{consensus+innovations} estimator proposed in~\cite{Ourwork1} that employs  general nonlinearities in the consensus and innovations update part, and we provide novel analysis and insights into its behavior.  
Employing the general nonlinearity in the presented method 
ensures provable robustness to heavy-tailed noise. In more detail, we present results on  almost sure (a.s.) convergence of the presented estimator to the true parameter~$\boldsymbol{\theta}^\star$. Moreover, asymptotic normality of the sequence of iterates generated by the algorithm is presented while also explicitly evaluating the corresponding asymptotic covariance.
%in terms of underlying system parameters -- the underlying network topology, the nonlinearity employed, and the observation and communication noise statistics.

With respect to~\cite{Ourwork1}, we provide here novel studies and insights. Namely,  we analyze the effect of the underlying network topology on the asymptotic variance of the estimator. The results reveal and quantify inherent tradeoffs between asymptotic variance and network topology. Intuitively, introducing more links in the network (``denser'' network topology) results in more communication noise injected overall, but it also leads to a stronger  ``useful information flow'' across the network. We explicitly quantify and illustrate the tradeoff on an analytical example, showcasing the existance of a nontrivial optimal tradeoff. 

%We further provide numerical results 
% confirming that, with the presented method,  the appropriately time-scaled  
% distance from the true parameter  (the distance multiplied by the square root of the time index~$t$) exhibits a ``compact distribution'' behavior with a bounded variance, thus confirming the asymptotic normality result. In contrast,  
%we show that, with a linear estimator~\cite{KMR}, the same quantity grows without bound.

%We now briefly review the literature to help us contrast further with existing work. 
%

There has been a large body of works on \emph{consensus +innovations} and related distributed estimation, e.g.,~\cite{A1,A4,E,G1,KMR,SoummyaAsymptEfficient,SoummyaAdaptive}, 
and distributed detection methods, e.g.,~\cite{SayedDetection,D1}. None of those works allows for presence of heavy-tailed noises. There have been several recent works that consider impulsive \emph{observation noise}, e.g.,~\cite{R1,R3,R5,prasad,F}. Among them, reference~\cite{F} employ certain nonlinearities in the innovation update part, in order to combat the impulsive observation noise. 
%This is in contrast with our work, that employs a nonlinearities in the \emph{consensus} and \emph{innovations} update part. 
Nonlinearities have been studied earlier in the context of standard average consensus problem~\cite{UsmanNolinConsesus,Stankovic,DasNonlinConsensus}. 
However, the methods studied in~\cite{UsmanNolinConsesus,Stankovic,DasNonlinConsensus} 
are very different from \emph{consensus+innovations}--the innovation update part does not exist. 
%Therefore, the analysis and results therein are very different from ours. 
In the context of distributed optimization, some specific nonlinearites have been considered in~\cite{Bianchi} and~\cite{Sundaram}. In contrast, 
we allow for a generic nonlinearity here (see ahead Assumption~\ref{as:2}).
%\textcolor{red}{In summary, with respect to existing work on consensus+innovations distributed inference, we employ for the first time a general nonlinearity in the consensus update, we allow for the first time for heavy-tail additive communication noise, and establish for the considered setting strong convergence guarantees, namely a.s. convergence and asymptotic normality.}

\textit{Notation}. We denote by $\mathbb R$ the set of real numbers and by ${\mathbb R}^m$ the $m$-dimensional
Euclidean space. We use normal lower-case letters for scalars,
lower case boldface letters for vectors, and upper case boldface letters for
matrices. Further,  to represent a vector $\mathbf{a}\in\mathbb{R}^m$ through its component, we write $\mathbf{a}=[\mathbf{a}_1, \mathbf{a}_2, ...,\mathbf{a}_m]^\top$ and we denote by: $\mathbf{a}_i$ or $[\mathbf{a}]_i$, as appropriate, the $i$-th element of vector $\mathbf{a}$; $\mathbf{A}_{ij}$ or $[\mathbf{A}]_{ij}$, as appropriate, the entry in the $i$-th row and $j$-th column of
a matrix $\mathbf{A}$;
$\mathbf{A}^\top$ the transpose of a matrix $\mathbf{A}$; $\otimes$ the Kronecker product of matrices. Further, we use either  
$\mathbf{a}^\top \mathbf{b}$ or 
$\langle \mathbf{a},\,\mathbf{b}\rangle$ 
for the inner products of vectors 
$\mathbf{a}$ and $\mathbf{b}$. Next, we let  
$\mathbf{I}$, $\mathbf{0}$, and $\mathbf{1}$ be, respectively, the identity matrix, the zero vector, and the column vector with unit entries. Further,  $\Diag(\mathbf{a})$ is the diagonal matrix 
whose diagonal entries are the elements of vector~$\mathbf{a}$; $\mathrm{Tr}(\mathbf{A})$ the trace of matrix $\mathbf{A}$; 
$\mathbf{J}$ the $N \times N$ matrix $\mathbf{J}:=(1/N)\mathbf{1}\mathbf{1}^\top$.
When appropriate, we indicate the matrix or vector dimension through a subscript.
We further denote by:
$\|\cdot\|=\|\cdot\|_2$ the Euclidean (respectively, spectral) norm of its vector (respectively, matrix) argument; $\lambda_i(\cdot)$ the $i$-th smallest eigenvalue;

\section{Model and Algorithm}
\label{section-model-algorithm}
%In the subsection~\ref{subsection-model}, considered network and observation 
%models are covered. In the subsection~\ref{subsection-algorithm-assumptions}
%we present the nonlinear consensus+innovations distributed estimator and also we state all necessary assumptions that are needed for the theoretical results presented in Section~\ref{section-main-results}.

%\subsection{Network, communication and observation models}

\subsection{Problem model}
\label{subsection-model}
We consider a problem where a network of $N$ agents aims to estimate an unknown (static) vector parameter $\boldsymbol{\theta}^{\ast}\in\mathbb{R}^{M}$.
At each time $t=0,1,...,$, each agent $i$ makes a scalar observation as follows:
\begin{align}
	\label{eq:obs_model}
	z_{i}^{t} = \mathbf{h}_{i}^{\top}\boldsymbol{\theta}^{\ast}+n_{i}^{t}.
\end{align}
Here, $z_{i}^{t}\in\mathbb{R}$ is the observation, $\mathbf{h}_{i}\in\mathbb{R}^{M}$ is the deterministic, non-zero linear transformation vector and $n_{i}^{t}\in\mathbb{R}$ is a scalar zero-mean noise.

\noindent The underlying network topology is defined through a graph $G=(V,E)$, where $V=\{1,...,N\}$  
is the set of agents, and $E$ is the set of (undirected) inter-agent communication links (edges)~$\{i,j\}$. 
We also let $\mathbf{L}$ denote the $N\times N$ (symmetric) graph Laplacian matrix, defined by $\mathbf{L}=\mathbf{D}-\mathbf{A}$, where $\mathbf{D}$ is the degree matrix and $\mathbf{A}$ is the adjacency matrix. That is, $\mathbf{D}=\Diag(\{d_i\})$, where $d_i$ is the degree (number of neighbors--excluding $i$) of agent $i$, and $\mathbf{A}$ has zero diagonal elements, while, for $i\neq j$, $\mathbf{A}_{ij}=1$ if and only if $\{i,j\}\in E.$
Next, we let $\Omega_i$ denote the neighborhood set of agent~$i$ (excluding~$i$). 
For an undirected edge $\{i,j\}\in E$, we denote by $(i,j)$ the arc that points from $j$ to $i$, and similarly, 
$(j,i)$ is the arc that points from $i$ to $j$.

\subsection{Algorithm and technical assumptions}
\label{subsection-algorithm-assumptions}
We now present the nonlinear \emph{consensus+innovations} estimator considered in this paper. 
The effect of the communication noises is alleviated by adding a general nonlinearity in the consensus part. In more detail,  
at each time $t=0,1,...$, each agent $i$ updates its estimate 
$\mathbf{x}_{i}^{t} \in {\mathbb R}^M$ of the parameter $\boldsymbol{\theta}^\ast$ via the following rule:
\begin{align}
	&\mathbf{x}_{i}^{t+1}=\mathbf{x}_{i}^{t}\nonumber\\&-\alpha_{t}\left(\frac{b}{a}\sum_{j\in\Omega_{i}}
	\boldsymbol{\Psi}_\mathrm{c}\left( \mathbf{x}_{i}^{t}-\mathbf{x}_{j}^{t} 
	+\boldsymbol{\xi}_{ij}^t\right)-\mathbf{h}_{i}{\Psi}_\mathrm{o}\left(z_{i}^{t}-\mathbf{h}_{i}^{\top}\mathbf{x}_{i}^{t}\right)\right).\label{eq:alg2}
\end{align}
Here, $\alpha_t = a/(t+1)$ is a step-size, 
$a,b>0$ are constants;  
$\boldsymbol{\xi}_{ij}^t \in {\mathbb R}^M$ is 
a zero-mean additive communication noise that models 
the imperfect communication from agent $j$ to agent $i$; $\boldsymbol{\Psi}_\mathrm{c}: \mathbb{R}^{M}\to\mathbb{R}^{M}$ is a non-linear map, given by:
\begin{align*}
	\boldsymbol{\Psi}_\mathrm{c}(\mathbf{y}_1,\mathbf{y}_2,...,\mathbf{y}_M)=[\Psi_\mathrm{c}(\mathbf{y}_1),\Psi_\mathrm{c}(\mathbf{y}_2),...,\Psi_\mathrm{c}(\mathbf{y}_M)]^\top,
\end{align*}
where 
$\Psi_\mathrm{c}: \, \mathbb{R} \to\mathbb{R} $ is a component-wise non-linear function.  
\noindent We now comment on algorithm~\eqref{eq:alg2}.
The innovation update part is given by the term ${\Psi}_\mathrm{o}\left(\mathbf{h}_{i}\left({z}_{i}^{t}-\mathbf{h}_{i}^{\top}\mathbf{x}_{i}^{t}\right)\right)$, as it assimilates 
the newly acquired observation $z_i^t$ corrupted by sensing heavy tailed noise. Moreover, the term $\sum_{j\in\Omega_{i}}
\Psi\left( \mathbf{x}_{i}^{t}-\mathbf{x}_{j}^{t} 
+\boldsymbol{\xi}_{ij}^t\right)$ in \eqref{eq:alg2} 
corresponds to consensus part, i.e., adapting 
the agent $i$'s estimate $\mathbf{x}_{i}^{t} $
by taking into account the noisy versions of the estimates of its neighbors 
$\mathbf{x}_{j}^{t} - \boldsymbol{\xi}_{ij}^t$, $j \in \Omega_i$, 
that agent $i$ receives at time~$t$. 
In conventional, linear \emph{consensus+innovations}, e.g., \cite{KMR}, 
the consensus and innovations terms corresponds respectively to 
a linear operation: $\sum_{j\in\Omega_{i}}
\left( \mathbf{x}_{i}^{t}-\mathbf{x}_{j}^{t} 
+\boldsymbol{\xi}_{ij}^t\right)$ and $\mathbf{h}_{i}\left({z}_{i}^{t}-\mathbf{h}_{i}^{\top}\mathbf{x}_{i}^{t}\right)$  i.e., function $\Psi$ equals 
indentity. In contrast, we allow $\Psi$ to be 
a general nonlinearity, 
see ahead Assumption~\ref{as:2}. Intuitively, 
the nonlinearity $\Psi$ usually has a ``saturation/truncation form'' (e.g., a clipping nonlinearity) that reduces the injection of noise in the iterates. 
This comes at a cost of also reducing the degree of inclusion of 
useful information from the neighbors. As demonstrated in Theorems~\ref{theorem-almost-surely} and~\ref{theorem-asymptotic-normality}, below, 
the net effect is positive in the presence of heavy-tail noise. 
In fact, setting $\Psi$ equal identity actually leads to 
infinite-variance solution estimates~\cite{Ourwork}.
\noindent We next specify the assumptions that we make on the underlying network, non-linear map, and the observation and communication noises.

\begin{myassump}\label{as:1}\textbf{Network model:}
	Graph $G=(V,E)$ is undirected, simple (no self nor multiple links), static and connected, i.e., $\lambda_{2}\left(\mathbf{L}\right)>0$.
\end{myassump}

\begin{myassump}\label{as:2}\textbf{Nonlinearity $\Psi$:}
	The non-linear function $\Psi:\mathbb{R}\to\mathbb{R}$ satisfies the following properties:
	1. Function~$\Psi$ is odd, i.e., $\Psi(a)=-\Psi(-a),$ for any $a\in\mathbb{R}$;
	2. $\Psi(a) > 0,$ for any $a > 0$.
	3. Function $\Psi$ is a monotonically nondecreasing function;
	4. $\Psi$ is continuous, except possibly on a point set with Lebesque measure of zero. Moreover, $\Psi$ is piecewise differentiable;
	5. $|\Psi(a)|\leq c_1$, for some constant $c_1>0.$
	6. $\Psi$ is either discontinuous at zero, or $\Psi(u)$ is strictly increasing for $u\in(-c_2,c_2)$, for some $c_2 > 0.$
\end{myassump}

There are many interesting examples of nonlinearities 
that fall within our framework, such as 
the sign or  clipping function, symmetric quantizers, etc.; see~\cite{Ourwork} for details.

\begin{myassump}\label{as:3}\textbf{Observation model:}
	1. For each agent 
	$i=1,...,N$, the observation noise sequence $\{{n}_{i}^{t}\}$ in \eqref{eq:obs_model},
	is independent identically distributed (i.i.d.);
	2. At each agent $i=1,...,N$~at each time~$t=0,1,...,$ noise $n_i^t$ has the same probability density function~$p_\mathrm{o}$.
	% \item At Each agent $i=1,...,N$~at each time~$t=0,1,...,$ noise $n_i^t$ has the same cumulative distribution function (cdf)~$\Phi_\mathrm{o}$.
	3. Random variables ${n}_{i}^{t}$ and ${n}_{j}^{s}$ 
	are mutually independent whenever the tuple  $(i,t)$ 
	is different from $(j,s)$;
	4. The pdf $p_\mathrm{o}$ is symmetric, i.e. $p_\mathrm{o}(u)=p_\mathrm{o}(-u),$ for every $u\in\mathbb{R}$, and $p_\mathrm{o}(u)>0$ for $|u|\leq c_\mathrm{o}$, for some constant $c_\mathrm{o}>0$;
	5. There holds that with $\int |u|p_\mathrm{o}(u)du <\infty$.
\end{myassump}

%Assumption \ref{as:3} is standard. Specifically, invertibility of matrix $\sum_{i=1}^{N}\mathbf{h}_i\mathbf{h}_i^{\top}$ is a necessary condition, even for a centralized estimator that collects observations according to~\eqref{eq:obs_model}, to be consistent.

\begin{myassump}\label{as:4} \textbf{Communication noise:}
	1. Additive communication noise $\{\boldsymbol{\xi}^t_{ij}\}$, $\boldsymbol{\xi}^t_{ij} \in \mathbb{R}^M$ is i.i.d. in time $t$, and independent across different arcs~$(i,j)\in E_d.$
	2. Each random variable $[\boldsymbol{\xi}^t_{ij}]_{\ell}$, for 
	each $t=0,1...$, for each arc $(i,j)$, 
	for each entry $\ell=1,...,M$, 
	has the same probability density function~$p_\mathrm{c}$.
	3. The pdf $p_\mathrm{c}$ is symmetric, i.e. $p_\mathrm{o}(u)=p_\mathrm{c}(-u),$ for every $u\in\mathbb{R}$ and $p_\mathrm{c}(u)>0$ for $|u|\leq c_\mathrm{c}$, for some constant $c_\mathrm{c}>0$;
	4. There holds that $\int |u|p_\mathrm{c}(u)du <\infty$.
\end{myassump}
As it can be seen, in the assumptions~\ref{as:3} and~\ref{as:4}, infinite variance of noises is allowed. Moreover, there is no assumption that communication noise and observation noise are mutually dependent.
For simplicity, we let the communication noise have the same distribution $\Phi$ for all arcs $(i,j)$, $\{i,j\}\in E$. Similarly, we let each element of communication noise vector $[\boldsymbol{\xi}^t_{ij}]_\ell$, $\ell=1,2,...,M$ have the same cumulative distribution function, and that $[\boldsymbol{\xi}^t_{ij}]_\ell$ and $[\boldsymbol{\xi}^t_{ij}]_s$ are mutually independent for $\ell\neq s$. For extensions to non-equal nonlinearities across links and non-equal, mutually dependent, communication noise distributions $[\boldsymbol{\xi}^t_{ij}]_\ell$ and $[\boldsymbol{\xi}^t_{ij}]_s$ for $\ell\neq s$, see~\cite{Ourwork}.
Furthermore, a more general assumptions set for the results to hold can be found in~\cite{Ourwork}.

\section{Main results}
\label{section-main-results}
In this section, we present theoretical results that are established under Assumptions~\ref{as:1}-\ref{as:4}. Firstly, we present the result on almost sure convergence of distributed estimator~\eqref{eq:alg2}. Then, in  Theorem~\ref{theorem-asymptotic-normality}, shows that \eqref{eq:alg2} is asymptotically normal, where we also evaluate the corresponding asymptotic variance. We have the following theorems, proofs of which can be found in \cite{Ourwork}.

\begin{theorem}[Almost sure convergence]
	\label{theorem-almost-surely}
	Let Assumptions \ref{as:1}-\ref{as:4} hold and $\alpha_t= a/(t+1)^\delta,$ $\delta\in (0.5,1]$.  
	Then, for each agent $i=1,...,N$, 
	the sequence of iterates $\{\mathbf{x}_i^t\}$ 
	generated by algorithm~\eqref{eq:alg2} 
	converges almost surely to the true 
	vector parameter~$\boldsymbol{\theta}^{\ast}$.
\end{theorem}

\begin{theorem}[Asymptotic normality]
	\label{theorem-asymptotic-normality}
	Let Assumptions \ref{as:1}-\ref{as:4} hold. 
	Consider algorithm~\eqref{eq:alg2} with step-size 
	$\alpha_t=a/(t+1)^{\delta}$, $t=0,1,...,$ $a>0$, with  $\delta=1$. Then, 
	the normalized sequence of iterates $\{
	\sqrt{t+1} (  \mathbf{x}^t-\mathbf{1}_N\otimes\boldsymbol{\theta}^\ast ) \}$ 
	converges in distribution to a zero-mean multivariate normal random vector, i.e., the following holds:
	\begin{align*}
		\sqrt{t+1}(\mathbf{x}^t-\mathbf{1}_N\otimes\boldsymbol{\theta}^\ast)\Rightarrow\mathcal{N}(\mathbf{0},\mathbf{S}),
	\end{align*}
	where the asymptotic covariance matrix $\mathbf{S}$ equals:
	\begin{equation}
		\label{eqn-asympt-var}
		\mathbf{S}=a^2\int\limits_{0}^\infty e^{\boldsymbol{\Sigma} v}\mathbf{S}_0e^{\boldsymbol{\Sigma}^\top v}dv.
	\end{equation}
	Here,  
	$\mathbf{S}_0=\frac{b^2}{a^2}\sigma_\mathrm{c}^2\Diag\left( \{d_i\,\mathbf{I}_M\}\right)-\frac{b}{a}\mathbf{K}_\mathrm{c,o} \mathbf{H} -\frac{b}{a}\mathbf{H}^\top\mathbf{K}_\mathrm{c,o}^\top  +\sigma_\mathrm{o}^2\mathbf{H}^\top\mathbf{H}$;
	$\sigma_\mathrm{o}^2 = \int |\Psi_\mathrm{o}(w)|^2 $ $d \Phi_\mathrm{o}(w)$
	is the effective observation noise variance 
	after passing through the nonlinearity~$\Psi_\mathrm{o}$;
	$\sigma_\mathrm{c}^2 = \int |\Psi_\mathrm{c}(w)|^2 d \Phi_\mathrm{c}(w)$
	is the effective communication noise variance 
	after passing through the nonlinearity~$\Psi_\mathrm{c}$; $\mathbf{K}_\mathrm{c,o}\in\mathbf{R}^{MN\times N}$ is the effective cross-covariance matrix between the observation and the communication noise after passing through the appropriate nonlinearity, i.e., the $(k,s)$ element of the matrix $\mathbf{K}_\mathrm{c,o}$ is given by $[(\mathbf{K}_\mathrm{c,o})]_{ks}=\sum\limits_{j\in\Omega_i}\int\int\Psi_\mathrm{c}(w_{ij\ell})\Psi_\mathrm{o}(w_k)p^{\mathrm{c,o}}_{k,ij\ell}(w_{ij\ell},w_k)dw_{ij\ell}dw_k.$ Here, $\ell$ satisfies the following: $s=M(i-1)+\ell$; and $p^{\mathrm{c,o}}_{k,ij\ell}$ is the joint probability density function for the $k$-th observation noise $n_k$ and the  $\ell$-th element of the communication noise $[(\boldsymbol{\xi}_{ij})]_\ell$.
	We also recall the observation matrix $\mathbf{H}$ in \eqref{eq:obs_model};  
	functions $\varphi_\mathrm{c}$, $\varphi_\mathrm{o}$ auxiliary nonlinearities (see~\cite{Ourwork1};  
	and $\Sigma=\frac{1}{2}\mathbf{I}-a(\frac{b}{a}\varphi_\mathrm{c}^\prime(0)\mathbf{L}\otimes \mathbf{I}_M + \varphi_\mathrm{o}^\prime(0) \mathbf{H}^{\top}\mathbf{H});$ here, $a$ is taken large enough such that matrix $\boldsymbol{\Sigma}$ is stable.
\end{theorem}
%     \begin{theorem}[MSE convergence]\label{theorem-MSE} Let Assumptions \ref{as:1}-\ref{as:4} hold. Then, for the sequence of iterates $\{\mathbf{x}^t\}$ 
	% 	generated by algorithm~\eqref{eq:alg2}, provided that the step-size sequence $\{\alpha_t\}$ is given by $\alpha_t=a/(t+1)^\delta,$ $a>0,\delta\in(0.5,1)$, there exists $\hat{\delta} \in (0,1)$ such that $\mathbb{E}[\|\mathbf{x}-\mathbf{1}_{N}\otimes \boldsymbol{\theta}^{\ast}\|^2]=O(1/t^{\hat{\delta}}).$ 
	% \end{theorem}
Theorem~\ref{theorem-almost-surely} and Theorem~\ref{theorem-asymptotic-normality} establish almost sure convergence and asymptotic normality
of estimator~\eqref{eq:alg2}. 
On the other hand, the linear \emph{consensus+innovations} scheme in \cite{KMR} (recovered by setting $\Psi$ to identity) leads to a sequence with unbounded second moments for all $t=1,2,...$, under the noise setting of Theorem 2 (see~\cite{Ourwork1}).

\section{Analytical example}
\label{section-examples}
\noindent In this section, we provide a study on the effect of network topology on the asymptotic variance of the estimator. We let parameter  $\theta^\ast\in\mathbb{R}$ be a scalar; for each agent $i$ at each time $t$, the  observation is given by  $z_i(t)=h\theta^\ast+n_i^t.$ Here, $h\neq 0$ is a deterministic parameter, and sequence $\{n_i^t\}$ satisfies Assumption~\ref{as:3}. We denote the effective observation noise variance after passing through the nonlinearity $\Psi_\mathrm{o}$ by $\sigma^2_{\textrm{o}}.$ We consider the communication noise that satisfies Assumption~\ref{as:4}, for which the effective variance is denoted by $\sigma^2_{\textrm{c}}$. We consider the nonlinearities $\Psi_{\mathrm{o}}(w)=\Psi_{\mathrm{c}}(w)=\sign{w}.$ We assume that the underlying graph is a regular graph with degree $d$.
Using Theorem~\ref{theorem-asymptotic-normality}, it can be shown that, the average per-node asymptotic variance, $\sigma_{d}^2=\frac{1}{N} \Tr(\mathbf{S})$, is given by
$\sigma_{d}^2=\frac{a^2h^2\sigma_{\textrm{o}}^2+b^2d\sigma_{\mathrm{c}}^2}{N\left(4ah^2f_{\mathrm{o}}(0)-1\right)}+\frac{a^2h^2\sigma_{\textrm{o}}^2+b^2d\sigma^2_{\textrm{c}}}{N}\sum\limits_{i=2}^N\frac{1}{4b\lambda_if_{\textrm{c}}(0)+4ah^2f_{\textrm{o}}(0)-1},$
where $\sigma_{\textrm{o}}^2=\sigma_{\textrm{c}}^2=1$ and $f_{\textrm{o}}(w)$ and $f_{\textrm{c}}(w)$ are the pdfs of the observation and communication noise, respectively~\cite{Ourwork1}.  

\begin{figure}
	\centering
	\includegraphics[height=3.5cm]{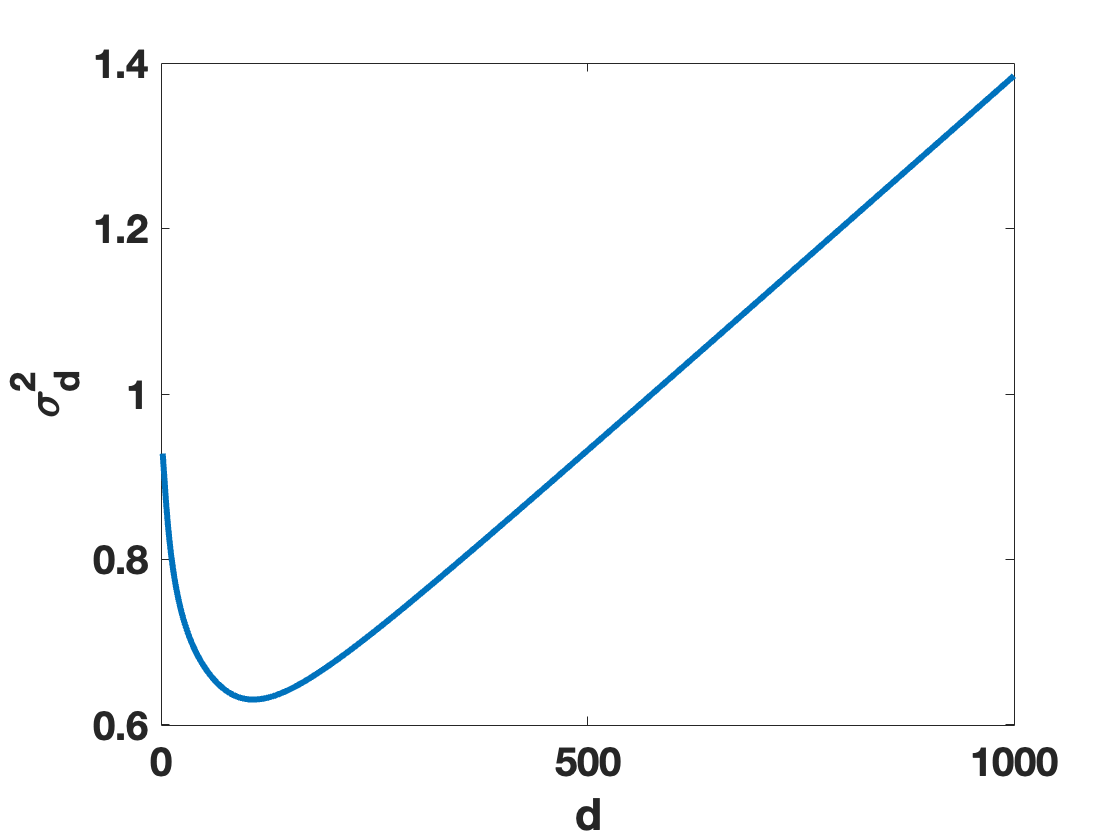}
	\caption{Per-agent asymptotic variance $\sigma_d^2$ versus $d$ for the nonlinear \emph{consensus+innovations} estimator and the $\Psi(w)=\sign (w)$ nonlinearity.}
	\label{fig:sigmad}
\end{figure}

Our focus here is to analyze the behavior of $\sigma_{d}^2$ when we change the underlying network topology. That is, we examine $\sigma_{d}^2$ when the nodes' degree in the underlying regular graph varies. 
To do that, we first generate a ring graph (with degree 2). Then, for each agent $i$,  we add to $\Omega_i$ the neighbors of agents $j\in \Omega_i$, excluding $i.$ That is, the new graph is of degree 4 where each agent is connected to its two-hop neighbors in the ring. Repeating the procedure, we generate a sequence of regular graphs 
with degrees $2,4,6,...,(N-1)$, where the last, $(N-1)$-degree graph in the sequence is the complete graph.  
Figure~\ref{fig:sigmad} shows per-agent asymptotic variance $\sigma^2_d$ versus $d$ for $N=1001$ agents, when the observation noise for each agent and the communication noise for each (directed) communication link has the following pdf:
\begin{align}\label{eqn:pdf}
	f_{\textrm{o}}(w)=f_{\textrm{c}}(w)=\frac{\beta-1}{2\,(1+|w|)^\beta}, 
\end{align}
with $\beta=2.05$. Here, we set $a=b=h=1.$ Note that, for each different graph topology, 
the communication noise considered \emph{per each directed link} is kept constant.

We can see in Figure~\ref{fig:sigmad} that adding more links induces a tradeoff with respect to asymptotic variance. Intuitively, on the one hand, adding more links injects more communication noise in the system overall. On the other hand, as number of links increases, the ``useful information flow'' also becomes faster across the network. Therefore, there is a degree value ($d=108$ for this numerical example) that balances the two effects and hence minimizes~$\sigma^2_d.$

\section{Conclusion}
\label{section-conclusion}
We presented a \emph{consensus+innovations} distributed estimator 
that employs a generic nonlinearity in the consensus and innovations update parts. 
We presented an analysis that 
shows almost sure convergence to the true unknown parameter and asymptotic normality of the 
nonlinear estimator, under a general setting for communication and observation noises that may have infinite variance. Under the same setting, traditional \emph{linear consensus+innovations} distributed estimators fail to converge. Novel analytical studies are provided that explain and quantify tradeoffs between the estimator's asymptotic variance, system noises, and the underlying network topology.

	\bibliographystyle{siam}
\bibliography{references.bib}

@ARTICLE{KMR,
	author={Kar, Soummya and Moura, José M. F. and Ramanan, Kavita},
	journal={IEEE Transactions on Information Theory}, 
	title={Distributed Parameter Estimation in Sensor Networks: Nonlinear Observation Models and Imperfect Communication}, 
	year={2012},
	volume={58},
	number={6},
	pages={3575-3605},
	doi={10.1109/TIT.2012.2191450}}

@article{SoummyaAsymptEfficient,
	author = {Kar, Soummya and Moura, Jose},
	year = {2014},
	month = {01},
	pages = {4811-4831},
	title = {Asymptotically Efficient Distributed Estimation With Exponential Family
	Statistics},
	volume = {60},
	journal = {IEEE Transactions on Information Theory},
	doi = {10.1109/TIT.2014.2331272}
}

@article{SoummyaAdaptive,
	author = {Kar, Soummya and Moura, Jose and Poor, H. Vincent},
	year = {2013},
	month = {09},
	pages={2200-2229},
	title = {Distributed Linear Parameter Estimation: Asymptotically Efficient
	Adaptive Strategies},
	volume = {51},
	journal = {SIAM Journal on Control and Optimization},
	doi = {10.1137/110848396}
}

@inproceedings{UsmanNolinConsesus,
	author={Khan, Usman A. and Kar, Soummya and Moura, José M. F.},
	booktitle={2009 Conference Record of the Forty-Third Asilomar Conference on Signals, Systems and Computers}, 
	title={Distributed average consensus: Beyond the realm of linearity}, 
	year={2009},
	volume={},
	number={},
	pages={1337-1342},
	doi={10.1109/ACSSC.2009.5469905}
}

@ARTICLE{DasNonlinConsensus,
  author={Dasarathan, Sivaraman and Tepedelenlioğlu, Cihan and Banavar, Mahesh K. and Spanias, Andreas},
  journal={IEEE Transactions on Signal Processing}, 
  title={Robust Consensus in the Presence of Impulsive Channel Noise}, 
  year={2015},
  volume={63},
  number={8},
  pages={2118-2129},
  doi={10.1109/TSP.2015.2408564}}

@inproceedings{Stankovic,
	author = {Stankovic, Srdjan and Beko, Marko and Stankovic, Milos},
	year = {2019},
	pages = {1-6},
	booktitle={IEEE EUROCON 2019-18th International Conference on Smart Technologies},
	title = {A Robust Consensus Seeking Algorithm},
	doi = {10.1109/EUROCON.2019.8861907}
}

@inproceedings{Sundaram,
	author = {Sundaram, Shreyas and Gharesifard, Bahman},
	year = {2015},
	booktitle={2015 53rd Annual Allerton Conference on Communication, Control, and Computing (Allerton)},
	pages = {244-249},
	title = {Consensus-based distributed optimization with malicious nodes},
	doi = {10.1109/ALLERTON.2015.7447011}
}

@article{Bianchi,
	author={Ben-Ameur, Walid and Bianchi, Pascal and Jakubowicz, Jérémie},
	journal={IEEE Transactions on Automatic Control}, 
	title={Robust Distributed Consensus Using Total Variation}, 
	year={2016},
	volume={61},
	number={6},
	pages={1550-1564},
	doi={10.1109/TAC.2015.2471755}
}

@article{IoTHeavyTail,
	 author={Clavier, Laurent and Pedersen, Troels and Larrad, Ignacio and Lauridsen, Mads and Egan, Malcolm},
	journal={IEEE Communications Letters}, 
	title="{Experimental evidence for heavy tailed interference in the IoT}", 
	year={2021},
	volume={25},
	number={3},
	pages={692-695},
	doi={10.1109/LCOMM.2020.3034430}}

@article{A1,
	 author={Lalitha, Anusha and Javidi, Tara and Sarwate, Anand D.},
	journal={IEEE Transactions on Information Theory}, 
	title={Social Learning and Distributed Hypothesis Testing}, 
	year={2018},
	volume={64},
	number={9},
	pages={6161-6179},
	doi={10.1109/TIT.2018.2837050}
}

@inproceedings{A4,
	title={Nonasymptotic convergence rates for cooperative learning over time-varying directed graphs},
	author={Nedic, Angelia and Olshevsky, Alex and Uribe, Cesar A},
	booktitle={2015 American Control Conference (ACC)},
	pages={5884--5889},
	year={2015},
	organization={IEEE},
	doi = {10.1109/ACC.2015.7172262}
}

@article{SayedDetection,
	author={Matta, Vincenzo and Braca, Paolo and Marano, Stefano and Sayed, Ali H.},
	journal={IEEE Transactions on Information Theory}, 
	title={Diffusion-Based Adaptive Distributed Detection: Steady-State Performance in the Slow Adaptation Regime}, 
	year={2016},
	volume={62},
	number={8},
	pages={4710-4732},
	doi={10.1109/TIT.2016.2580665}
}

@article{D1,
	author = {Bajovic, Dragana and Jakovetic, Dusan and Xavier, Joao and Sinopoli, Bruno and Moura, Jose},
	year = {2011},
	month = {10},
	pages = {4381 - 4396},
	title = {Distributed Detection via Gaussian Running Consensus: Large Deviations Asymptotic Analysis},
	volume = {59},
	journal = {Signal Processing, IEEE Transactions on},
	doi = {10.1109/TSP.2011.2157147}
}

@article{B,
	author = {Cao, Ming and Morse, A. and Anderson, Brian},
	year = {2008},
	month = {01},
	pages = {575-600},
	title = {Reaching a Consensus in a Dynamically Changing Environment: A Graphical Approach},
	volume = {47},
	journal = {SIAM J. Control and Optimization},
	doi = {10.1137/060657005}
}

@inproceedings{C,
	author={Fagnani, Fabio and Zampieri, Sandro},
	booktitle={Proceedings of the 45th IEEE Conference on Decision and Control}, 
	title={Average consensus with packet drop communication}, 
	year={2006},
	volume={},
	number={},
	pages={1007-1012},
	doi={10.1109/CDC.2006.377319}
}

@article{D,
	author = {Huang, Minyi and Manton, Jonathan},
	year = {2009},
	month = {01},
	pages = {134-161},
	title = {Coordination and Consensus of Networked Agents with Noisy Measurements: Stochastic Algorithms and Asymptotic Behavior},
	volume = {48},
	journal = {SIAM J. Control and Optimization},
	doi = {10.1137/06067359X}
}

@article{E,
	author = {Mateos, Gonzalo and Schizas, Ioannis and Giannakis, G.B.},
	year = {2009},
	month = {12},
	pages = {4583 - 4588},
	title = {Distributed Recursive Least-Squares for Consensus-Based In-Network Adaptive Estimation},
	volume = {57},
	journal = {Signal Processing, IEEE Transactions on},
	doi = {10.1109/TSP.2009.2024278}
}

@article{F,
	author={Al-Sayed, Sara and Zoubir, Abdelhak M. and Sayed, Ali H.},
	journal={IEEE Transactions on Signal Processing}, 
	title={Robust Distributed Estimation by Networked Agents}, 
	year={2017},
	volume={65},
	number={15},
	pages={3909-3921},
	doi={10.1109/TSP.2017.2703664}
}

@article{I,
	author = {Theodoridis, S. and Slavakis, Konstantinos and Yamada, Isao},
	year = {2011},
	month = {02},
	pages = {97 - 123},
	title = {Adaptive Learning in a World of Projections},
	volume = {28},
	journal = {Signal Processing Magazine, IEEE},
	doi = {10.1109/MSP.2010.938752}
}

@article{R1,
	author = {Modalavalasa, Sowjanya and Sahoo, Upendra and Sahoo, Ajit and Baraha, Satyakam},
	year = {2021},
	month = {05},
	pages = {108150},
	title = {A Review of Robust Distributed Estimation Strategies Over Wireless Sensor Networks},
	volume = {188},
	journal = {Signal Processing},
	doi = {10.1016/j.sigpro.2021.108150}
}

@article{R3,
	author = {Li, Zhi and Guan, Sihai},
	year = {2018},
	month = {03},
	pages={3812--3825},
	title = {Diffusion normalized Huber adaptive filtering algorithm},
	volume = {355},
	journal = {Journal of the Franklin Institute},
	doi = {10.1016/j.jfranklin.2018.03.001}
}

@article{R5,
	author = {Wen, Fuxi},
	year = {2013},
	month = {07},
	pages = {},
	title = {Diffusion Least Mean P-Power Algorithms for Distributed Estimation in Alpha-Stable Noise Environments},
	volume = {49},
	journal = {Electronics Letters},
	doi = {10.1049/el.2013.2331}
}

@ARTICLE{G1,
	author={Zhao, Xiaochuan and Tu, Sheng-Yuan and Sayed, Ali H.},
	journal={IEEE Transactions on Signal Processing}, 
	title={Diffusion Adaptation Over Networks Under Imperfect Information Exchange and Non-Stationary Data}, 
	year={2012},
	volume={60},
	number={7},
	pages={3460-3475},
	doi={10.1109/TSP.2012.2192928}
}

@article{prasad,
	title={Robust estimation via robust gradient estimation},
	author={Prasad, Adarsh and Suggala, Arun Sai and Balakrishnan, Sivaraman and Ravikumar, Pradeep},
	journal={Journal of the Royal Statistical Society: Series B (Statistical Methodology)},
	volume={82},
	number={3},
	pages={601--627},
	year={2020},
	publisher={Wiley Online Library}
}

@article{Haenggi,
author = {Haenggi, Martin and Ganti, Radha},
year = {2009},
month = {01},
pages = {127-248},
title = {Interference in Large Wireless Networks},
volume = {3},
journal = {Foundations and Trends in Networking},
doi = {10.1561/1300000015}
}

@article{Ourwork,
	author = {Jakovetic, Dusan and Vukovic, Manojlo and Bajovic, Dragana and Sahu, Anit Kumar and Kar, Soummya},
title = {Distributed Recursive Estimation under Heavy-Tail Communication Noise},
journal = {SIAM Journal on Control and Optimization},
volume = {61},
number = {3},
pages = {1582-1609},
year = {2023},
doi = {10.1137/22M1477015},

URL = { 
    
        https://doi.org/10.1137/22M1477015
    
    

},
eprint = { 
    
        https://doi.org/10.1137/22M1477015
    
    

}
}

@article{Ourwork1,
author = {Vukovic, Manojlo and Jakovetic, Dusan and Bajovic, Dragana and Kar, Soummya},
title = {Nonlinear Consensus+Innovations under Correlated Heavy-Tailed Noises: Mean Square Convergence Rate and Asymptotics},
journal = {SIAM Journal on Control and Optimization},
volume = {62},
number = {1},
pages = {376-399},
year = {2024},
doi = {10.1137/22M1543197},

URL = { 
    
        https://doi.org/10.1137/22M1543197
    
    

},
eprint = { 
    
        https://doi.org/10.1137/22M1543197
    
    

}
}

\end{document}